\documentclass[preprint]{elsarticle}
\biboptions{sort&compress}
\usepackage{graphics}
\usepackage{amssymb}

\journal{Solid State Communications}

\begin{document}

\begin{frontmatter}

\title{Homogeneity of Bilayer Graphene}

\author[unibas]{Frank Freitag}

\author[unibas]{Markus Weiss\corref{cor1}}
\ead{Markus.Weiss@unibas.ch}

\author[unibas]{Romain Maurand}

\author[unibas]{Jelena Trbovic}

\author[unibas]{Christian Sch\"onenberger}

\address[unibas]{Department of Physics, University of Basel, Klingelbergstr. 82, CH-4056 Basel, Switzerland}
\cortext[cor1]{Corresponding author}

\begin{abstract}
We present non-linear transport measurements on suspended, current
annealed bilayer graphene devices. Using a multi-terminal geometry
we demonstrate that devices tend to be inhomogeneous and host two
different electronic phases next to each other. Both of these phases
show gap-like features of different magnitude in non-linear
transport at low charge carrier densities, as already observed in
previous studies. Here, we investigate the magnetic field dependence
and find that both features grow with increasing field, the smaller
one with 0.6~meV/T, the larger one with a 5-10 times higher
field dependence. We attribute the larger of the two gaps to an
interaction induced broken symmetry state and the smaller one to
localization in the more disordered parts of the device.
\end{abstract}


\end{frontmatter}

\section{Introduction}
\label{Introduction} The isolation of monolayer
graphene \cite{Novoselov2004} has started a new research field on
two-dimensional carbon based electronic systems. Whereas the
interest in monolayer graphene mainly stems from its unusual
pseudo-relativistic electron dispersion, bilayer graphene promises
to be interesting because of its strongly interacting charge
carriers. Weak screening in suspended samples and small carrier
densities at the charge neutrality point lead to large values for
the interaction parameter r$_S$ about one order of magnitude larger
than in conventional 2DEGs of comparable carrier density
\cite{DasSarma2011}. For interaction effects to be observable in
electronic transport experiments, bilayer graphene has to be very
clean to achieve low charge carrier densities. To reach this regime,
the normally used SiO$_2$ substrate has to be removed by
under\-etching, leaving the bilayer freely suspended between the
contacts \citep{Bolotin2008,Feldman2009}. The suspension allows to
current anneal the bilayer graphene flake by Joule heating induced
by passing a large current through it. During this heating cycle
adsorbates may (at least partially) evaporate.

The exact nature of the electronic groundstate of bilayer graphene
has been investigated theoretically for several years
\cite{McCann2006b,Nilsson2006,Ezawa2007,Wang2007,Barlas2008,Min2008,Nandkishore2010b,Zhang2011, Rossi2011,Kharitonov2012,Lemonik2012}.
The fourfold degeneracy that stems from the spin and valley degrees
of freedom of the charge carriers will be broken by interactions,
and new ground\-states with partly or totally lifted degeneracies
are expected to occur. Interaction induced broken-symmetry states
that are being discussed as possible candidates for the groundstate
of bilayer graphene include the quantum anomalous Hall insulator
(QAH) \citep{Nandkishore2010b,Zhang2011}, the quantum spin Hall
insulator (QSH) \citep{Zhang2011}, the nematic phase
\citep{Vafek2010,Lemonik2010}, the layer antiferromagnetic phase
(LAF) \citep{Zhang2011,Kharitonov2011,Kharitonov2012}, and possibly
more. Most, but not all of these states show a gap in the excitation
spectrum around $E_F$=0, leading to a vanishing conductivity around
the charge neutrality point (CNP).

We can probe the electronic properties of the graphene bilayer as a
function of temperature, charge carrier density and magnetic field
by measuring the differential conductance $dI/dV=G_d$ as a function
of a bias voltage $V_{sd}$. We investigate low conducting and
insulating samples with similar features as in a previous study,
that exhibit two distinct gap-like features in
$G_{d}(V_{sd})$\cite{Freitag2012}, and that mainly differ in the
minimum conductance at $V_{sd}=0$. Here, we further investigate
these kinds of samples by looking at the magnetic field dependence of
$G_d(V_{sd})$ \cite{Velasco2011}. Our results suggest that the
samples are inhomogeneous after current annealing and are composed
of two different electronic phases. A possible reason for this is
that the process of current annealing does not clean the samples in
a uniform way, but creates almost disorder free areas in the center
of the samples and leaves some disorder at the edges, which leads to
a coexistence of electronic phases next to each other.

\section{Material and methods}
\label{MM}

Bilayer graphene was deposited onto a p$^+$-doped silicon wafer
covered with 300~nm silicon dioxide by micromechanical cleavage of
natural graphite. After deposition, a thin (5~nm) film of aluminium
was evaporated over the whole wafer. After exposure to air, this
film oxidised and transformed to Al$_2$O$_3$ almost completely.
Using standard electron-beam lithography, a PMMA etchmask was then
structured for shaping the graphene flake using an  Ar/O$_2$ plasma,
and in a second lithography step, the electrode structures were
defined. After each resist development, the exposed part of the
oxidised aluminium film was removed in a 25\%
tetra\-methyl\-ammonium hydroxide (TMAH) solution, thereby
eliminating any organic residues left on the graphene. The electrode
structures (3~nm Cr and 70~nm Au) were e-gun evaporated, and
subsequently the graphene devices were annealed in high vacuum at
10$^{-7}$~mbar and 200~$^\circ$C for several hours. The silicon
dioxide was then partly removed in 2\% buffered hydrofluoric acid,
leaving the graphene suspended 160~nm above the substrate. The
remaining silicon wafer acted as a gate with a voltage $V_g$
applied. The devices were mounted into a $^3$He cryostat in vacuum
and subsequently current annealed at $T=1.5$ K with typically 0.4~mA
per $\mu$m of flake width \citep{Freitag2012}. For the present
study, we fabricated devices with a four-terminal Hall cross
geometry (insets of fig. \ref{fig1}); the longer axis was about
3~$\mu$m long, the shorter 1.5~$\mu$m and the width of each
individual arm was 1~$\mu$m.
Current annealing was done on pairs of contacts with the other two
contacts floating. The annealing process was considered to be
successful if a pronounced dip in $G$ measured as a function of the
gate-voltage $V_g$ appeared in the vicinity of $V_g=0$, i.e. at low
charge carrier density of at least one contact pair combination.

\section{Results}
\label{results}

\begin{figure}[htb]
\centering
\includegraphics{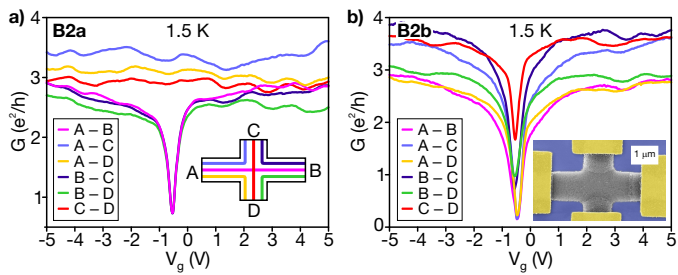}
\caption{(a) Two-terminal conductance $G$ of a Hall-cross device as
function of the gate voltage $V_g$ at 1.5~K in device B2a. The
possible configuration are shown in the inset. Measurements
involving contact \textsl{B} show a pronounced dip in $G$ at $V_g
\approx$~-0.5~V. The remaining contact pairs have a flat gate
response. (b) $G$ as a function of $V_g$ for all possible combinations
of contacts in device B2b at 1.5~K. All curves show a minimum in $G$
at $V_g \approx$~-0.5~V. The inset shows a scanning electron
micrograph of the device.} \label{fig1}
\end{figure}

We present data taken on two devices that could successfully be
current  annealed. Fig. \ref{fig1} summarises our findings. In the
first device, B2a, the current annealing resulted in an
inhomogeneous cleaning of the graphene. Fig. \ref{fig1}a)
demonstrates that only in two-terminal measurements of $G(V_g)$ that
involve contact B a CNP is visible with a minimum conductance of
0.9~e$^2$/h. For all other contact pairs G only fluctuates around
3~e$^2$/h as $V_g$ is changed and does not display a notable
minimum. Further annealing rendered the sample more inhomogeneous,
leading to gate characteristics with several CNPs at different
$V_g$. In contrast, in B2b the current annealing yielded a more
homogeneous device. Fig. \ref{fig1}b) shows $G$ as function of $V_g$
for B2b. A CNP at $V_g \approx$~-0.5~V is well developed for all
contact pairs. A minimum of $G \approx$~0.2~e$^2$/h at the CNP  can
be found in $G(V_g)$ measured between contacts A and B. At higher
doping, the conductance saturates at 3-4~e$^2$/h.

\begin{figure}[htb]
\centering
\includegraphics{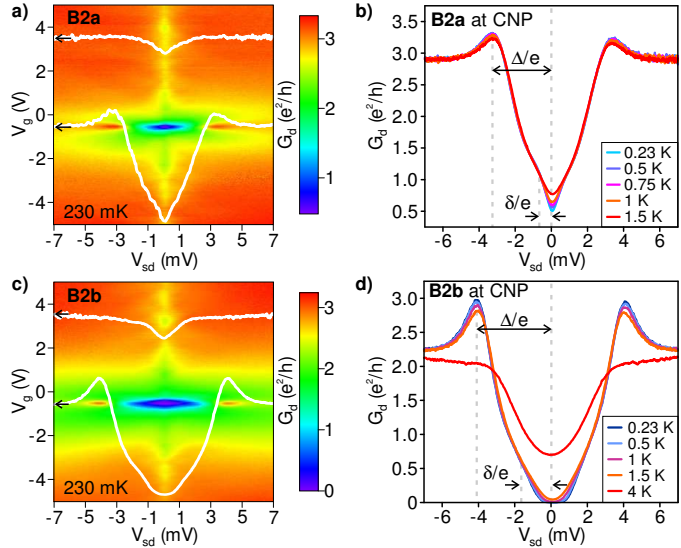}
\caption{ (a) Differential conductance $G_d$ as a function of gate
voltage  $V_g$ and bias voltage $V_{sd}$ for device B2a at
230~mK. Curves overlaid in white are $G_d(V_{sd})$-traces taken at
gate voltages indicated by the black arrows.
Close to the CNP at $V_g \approx$~-0.5~V $G_d$ is suppressed to below 1~e$^2$/h
for small bias voltages. At $|V_{sd}|=$~3.5~mV $G_d$ reaches a maximum
and then saturates around 3~e$^2$/h.
Away from the CNP, only a weak dependence on $V_{sd}$ is
observable. 
(b) Temperature dependence of $G_d$ as a function of
$V_{sd}$ of device B2a at the CNP. Two distinct gap features can be
identified. The smaller one, $\delta/e \approx$~0.6~mV, is reduced
when $T$ is raised from 0.23~K to 1.5~K. The larger gap, $\Delta / e
\approx$~3.5~mV, remains unchanged. (c) $G_d$ as a function of $V_g$
and $V_{sd}$ for device B2b at 230~mK. At the CNP, $V_g
\approx$~-0.5~V, and around zero $V_{sd}$ the device is insulating.
Changing $V_g$ or applying $V_{sd}$ recovers $G_d
\approx$~3~e$^2$/h. (d) Temperature dependence of $G_d$ as a function
of $V_{sd}$ at the CNP of B2b. The smaller gap, $\delta /e
\approx$~1.5~mV, is reduced when $T$ is increased from 0.23~K to
1.5~K. The larger gap, $\Delta /e \approx$~4~mV, is only reduced at
4~K.} \label{fig2}
\end{figure}

Performing differential conductance measurements $G_d(V_{sd})$
reveals  further differences between B2a and B2b. In fig.
\ref{fig2}a), a colour scale plot of $G_d$ in B2a as a function of
$V_g$ and $V_{sd}$ at 230~mK is shown. The line profile at the CNP
at $V_g \approx$~-0.5~V exhibits two gap-like features: a smaller
one, called $\delta$ in the following, is present up to $|V_{sd}|
=$~0.6~mV and reduces $G_d$ from 1~e$^2$/h to 0.5~e$^2$/h at zero
$V_{sd}$. The larger feature $\Delta$ is visible up to $|V_{sd}|
=$~3.5~mV and lowers $G_d$ from 3~e$^2$/h to about 1~e$^2$/h at zero
$V_{sd}$. Moreover, $\Delta$ shows a BCS-like overshoot in $G_d$.
Moving $V_g$ away from the CNP rapidly suppresses $\Delta$. On the
other hand, $\delta$ persists as a small zero bias anomaly even in
the metallic regime and is visible in the whole accessible gate
voltage range. In fig. \ref{fig2}b) the temperature dependence of
$G_d$ at the CNP of B2a is plotted. The smaller gap feature,
$\delta$, is totally suppressed if $T$ increases from 0.23~K to
1.5~K. The larger feature $\Delta$ remains unchanged.
The temperature dependence G(T) measured at large carrier density can be 
fit with a log(T) dependence for T$\leq$1K.

Device B2b differs from B2a by having a full gap, which reduces $G$
to zero at the CNP. Fig.
\ref{fig2}c) shows $G_d$ as a function of $V_g$ and $V_{sd}$ at
230~mK. The line profile at the CNP is qualitatively similar to B2a,
but $G_d$ is completely suppressed for small $V_{sd}$. Furthermore,
the small gap feature $\delta$ is not readily visible but manifests only as a
change in the slope of $G_d$ at $|V_{sd}| =$~1.5~mV. The larger
feature $\Delta$ is shifted to higher voltages as well, with $|V_{sd}|
=$~4~mV, and shows a larger overshoot before $G_d$ saturates at
2.3~e$^2$/h. Tuning $V_g$ away from the CNP quenches the larger
feature $\Delta$ but leaves the smaller $\delta$ intact in agreement
to what is observed in B2a. In fig. \ref{fig2}d) the temperature
dependence makes $\delta$ better visible. The small gap feature
shrinks as $T$ is raised from 0.23~K to 1.5~K, but still reduces
$G_d$ to zero at $V_{sd}=$~0~mV. A further increase to 4~K
completely suppresses $\delta$. The temperature change from 0.23~K
to 1.5~K only reduces the overshoot in the larger feature $\Delta$.
Increasing $T$ to 4~K yields a reduction of $\Delta$ and removes the
overshoot.

\begin{figure}[htb]
\centering
\includegraphics{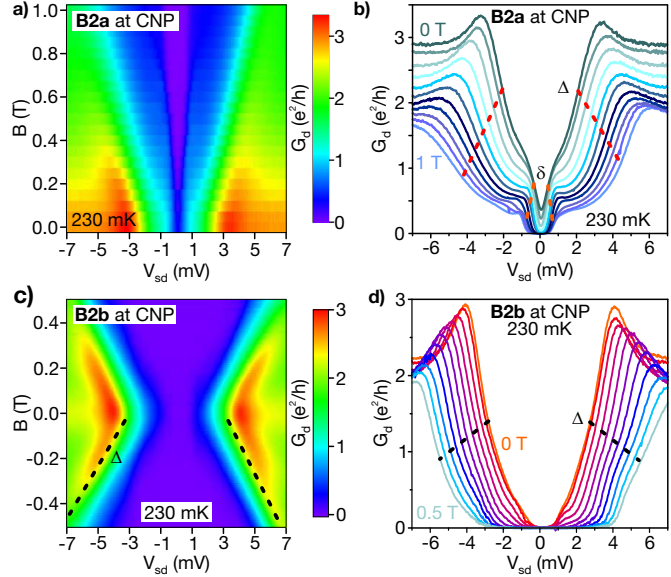}
\caption{ (a) Two-terminal conductance $G$ as a function of bias
voltage $V_{sd}$  and perpendicular magnetic field $B$ of device B2a
at the CNP and 230~mK. The device is low-conducting at $B=$~0~T and
smoothly evolves into an insulating state for $B>$~0.5~T. (b) The
corresponding line profiles. The dashed lines are guides to the eye
for the linear scaling of $\Delta$ and $\delta$ with $B$. $\Delta$
scales as 2.8~meV/T, whereas $\delta$ grows with 0.6~meV/T. (c) $G$
as a function of $V_{sd}$ and $B$ of device B2b at the CNP and 230~mK.
The device is insulating even at $B=$~0~T. The gap $\Delta$
increases linearly with $B$. (d) The corresponding line profiles.
The dashed line is a guide to the eye for the evolution of $\Delta$
with 6~meV/T.} \label{fig3}
\end{figure}

To find out about the origin of the two different features $\delta$
and $\Delta$ we perform differential conductance measurements
$G_d(V_{sd})$ as a function of perpendicular magnetic field, shown in figure
\ref{fig3}. Both $\delta$ and $\Delta$ increase linearly with $|B|$
but with different slope. Taking the two inflection points of the
$G_d(V_{sd})$ traces as a measure for $\delta$ and $\Delta$, we can
determine $\partial \delta / \partial B$ to 0.6 meV/T, which is
considerably larger than the Zeeman energy of a spin 1/2 (0.12 meV/T
for g=2). Looking at the large gap $\Delta$, we find $\partial
\Delta / \partial B$ = 2.8 meV/T for sample B2a (Fig. \ref{fig3}b)
and 6 meV/T for sample B2b (Fig \ref{fig3}d). As sample B2b showed a
full gap already at $B=0$, the magnetic field dependence of $\delta$
could not be determined here. Similarly, Velasco et al. observed
5.5~meV/T \citep{Velasco2011} for a fully gapped bilayer graphene
device for a gap of similar size as $\Delta$.

\begin{figure}[htb]
\centering
\includegraphics{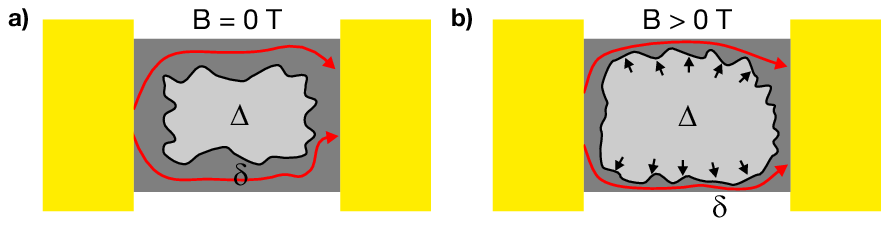}
\caption{ Sketch of the graphene device with the clean (light grey)
and  the disordered regions (dark grey) at the charge neutrality
point. (a) In the absence of an external magnetic field, the clean
region is gapped with $\Delta$ and the contacts do not couple to it.
The disordered region remains conductive (red arrows), but
experiences a transport gap $\delta$ due to localization. (b) In the
presence of a perpendicular magnetic field the clean region expands,
leading to a narrowing of the conducting areas $\delta$ and to a
suppression of conductance. } \label{fig4}
\end{figure}

The experiments clearly show two distinct gap-like features that
occur in different ranges of charge carrier density and temperature,
and that show a clearly different magnetic field dependence. We
think that this observation can only be explained by the coexistence
of different electronic phases next to each other in different parts
of the samples. This conjecture is supported by the results of the
current annealing procedures shown in figure \ref{fig1}, that can be
understood by the existence of almost disorder-free, clean regions
that are surrounded by more disordered areas \cite{DasSarma2010,Hwang2010}, where the current
annealing has been less efficient. In the disordered regions the CNP
lies outside the charge carrier densities that are accessible to us,
due to chemical doping by adsorbates and residues.
In this picture, sample B2a would be composed of a clean region
close to contact B, and a more disordered rest of the sample. In
sample B2b the clean region  seems to be larger, occupying the
center of the sample, but different CNP
conductances between different terminals indicate that even in
sample B2b there are regions with stronger disorder that are not
evenly distributed, as illustrated in Fig. \ref{fig4}.

We assume that the clean phase is showing an interaction induced gap
$\Delta$ close to the CNP with vanishing conductivity, and a
distinct, BCS like shape. The large increase of $\Delta$
with $B$ of several meV per tesla has been predicted by theory for
certain broken symmetry states like the QAH or QSH
\citep{Zhang2011,Velasco2011}.

The total conductance of sample B2a is composed of two parallel
conduction channels, one through the clean region with gap $\Delta$
in the center of the sample, and another, more disordered one along
the sample edges with gap-like feature $\delta$. The origin of the
smaller feature $\delta$ is probably disorder induced localization
\cite{Gorbachev2007,Oostinga2010,Poumirol2010}.
For a disordered device such as a graphene ribbon with a
predetermined geometry the application of a magnetic flux through the
sample should lift the localization at least in part. In contrast to
this scenario, the small gap in our experiment even increases when a
perpendicular magnetic field is applied. We explain this by a change
in geometry. The application of a perpendicular magnetic field
increases the size of the inner clean region and thereby narrows
down the conduction channel at the sample edge (Fig. \ref{fig4}b),
leading to increased localization in the disordered part. The
different response of the two conduction channels to magnetic field
is reflected in the different magnetic field dependence of $\partial
\delta / \partial B$ =0.6 meV/T and $\partial \Delta /\partial
B$=2.8 meV/T.

In sample B2b the initial distribution of the two phases is such
that already at B=0 the sample is mostly in the clean phase, and
there is only a negligible conductance contribution of the
disordered phase at the CNP. This is also the reason for the fact
that only a large gap $\Delta$ can be observed in $G_d(V_{sd})$, and
that $\delta$ is feeble.

Because the metallic contacts will act as a heat sink during current
annealing, we also have to assume that the disordered phase persists
close to them. In fact, a narrow area of organic deposits can be
clearly identified close to the metal contacts in scanning electron
micrographs. This disordered region will then decouple the clean
area in the sample center from the contacts and thus prevent edge
states, that exist in the case of a QAH or QSH phase to couple to
the metal contacts. Consequently, any phase in the center of the
sample would show insulating behaviour in two terminal conductance,
no matter if it has edge-states (QAH, QSH) or not (LAF). As we have
shown the current annealing process to be inhomogeneous, we think
that $\delta$ is a feature of the disordered graphene regions that
persist at the edges of the device. However, as the size of the
disordered region will vary from sample to sample, any distribution
of $\delta$ versus $\Delta$ -gaps may be realized. Different
distributions of clean and disordered regions in the samples will
lead to different $G_d(V_{sd})$ traces around the CNP, with a full
suppression of conductance at $V_{sd}$=0 when the clean region
dominates, and some small, finite conductance when there is a
considerable region of the disordered phase remaining. The
application of a perpendicular magnetic field will always lead to a
growth of the gapped $\Delta$-phase at the expense of the conducting
$\delta$-phase, in  this way opening a full gap in all samples at
some finite value of B. We note that, although not discussed in
detail, a small gap $\delta$ was also observed in
\citep{Velasco2011}. It persists at high charge carrier densities
and shows a small magnetic field dependence, in contrast to the
stronger magnetic field scaling of $\Delta$ and in full agreement to our observations.

\section{Conclusions}

In conclusion, we measured non-linear conductance on multi-terminal,
suspended bilayer graphene devices in a Hall-cross geometry as a
function of charge carrier density and perpendicular magnetic field.
We find that current annealing cleans the graphene inhomogeneously,
which leads to samples that are composed of two different regions
with distinct electronic properties. Both of these regions show
gap-like features at the CNP, as visible in differential conductance
$G_d(V_{sd})$. We extract two gaps of different magnitude that
show a different temperature dependence  and respond differently to
magnetic field. Attributing the two gaps to clean and disordered
sample regions, we find that the initial distribution of the two
regions differs from sample to sample, but that the overall density
and magnetic field dependence of the conductance of these two
regions is similar. We argue that one of the two regions is a
conductive, disordered phase localized at the sample edges that
shows a zero bias anomaly over a large range in charge carrier
density due to disorder induced localization, and a clean phase in
the center of the sample that shows an interaction induced gap close
to the charge neutrality point with strongly suppressed conductance.
The application of a perpendicular magnetic field favors the
expansion of the clean phase in the sample center and shrinks the
disordered conducting areas, and therefore pronounces the signature
of the interaction effects in differential conductance.

\section{Acknowledgments}
This work was financed by the Swiss NSF, the ESF programme Eurographene, the EU FP7 project SE$^2$ND, the Swiss NCCR Nano and QSIT. We are grateful to M. Kharitonov and C. N. Lau for discussions.

\end{document}